\documentclass[onecolumn]{aa}

\usepackage{graphicx}
\usepackage{txfonts}
\usepackage{color}

\begin{document}

   \title{Flare-related plasma motions in the outer atmosphere of the RS~CVn-type star II~Peg}

   \author{Dongtao Cao
          \inst{1,2,3}
          \and
          Shenghong Gu\inst{1,2,4}
          }

   \institute{Yunnan Observatories, Chinese Academy of Sciences, Kunming 650216, China\\
              \email{dtcao@ynao.ac.cn, shenghonggu@ynao.ac.cn}
         \and
             Key Laboratory for the Structure and Evolution of Celestial Objects, Chinese Academy of Sciences, Kunming 650216, China
         \and
             International Centre of Supernovae, Yunnan Key Laboratory, Kunming 650216, China
         \and
             School of Astronomy and Space Science, University of Chinese Academy of Sciences, Beijing 101408, China
             }
   \authorrunning{Cao \& Gu}
   \date{}

 
  \abstract
{Analogous to solar flares, stellar flares are dramatic explosions in the atmosphere, which may be accompanied by prominence eruptions, coronal mass ejections (CMEs), and other forms of plasma motion. Based on time-resolved spectroscopic observations of the RS CVn-type star II~Peg, we aim to search for the potential plasma motions associated with flares. In these observations, we detected part of the gradual decay phase of an optical flare, for which we find a lower limit on the energy of the H${\alpha}$ line of $6.03~\times~10^{33}$~erg. Converting this H${\alpha}$ energy, we find a bolometric white-light energy of $3.10~\times~10^{35}$~erg. Moreover, a secondary peak is also observed. After removing a quiescence reference, the H${\alpha}$ residual shows an asymmetric behavior, including both a blueshifted and a redshifted emission component. The former component has a bulk velocity of about -180~km~s$^{-1}$ and extends its velocity to more than -350~km~s$^{-1}$. This phenomenon is likely caused by a prominence eruption event or a chromospheric evaporation process. The latter emission component has a bulk velocity of 130 to 70~km~s$^{-1}$ and extends its velocity to nearly 400~km~s$^{-1}$. We attribute the redshifted emission component to one or a combination of several possible scenarios: flare-driven coronal rain, chromospheric condensation, backward-directed prominence eruption close to the stellar limb, or falling material in a prominence eruption. The minimum masses of the moving plasmas resulting in the blueshifted and redshifted emission components are estimated to be $0.56 \times 10^{20}$~g and $1.74 \times 10^{20}$~g, respectively.}

   \keywords{Stellar activity --
                Optical flares --
                Stellar coronal mass ejections --
                Stellar chromospheres --
                Spectroscopy
               }

   \maketitle
   \nolinenumbers
%
\section{Introduction}\label{sec1}
Solar flares are dramatic explosion events in the outer atmosphere of the Sun. They can be accompanied by prominence eruptions and coronal mass ejections (CMEs), which have significant influences on space weather \citep[e.g.,][]{Forbes2000, Koleva2021}. Although flares and even superflares are frequently observed on late-type stars \citep[e.g.,][]{Maehara2012, Tu2020, Tu2021, Yang2023}, particularly in the surveys of the {\it Kepler} space telescope and the Transiting Exoplanet Survey Satellite (TESS), reported stellar prominence eruption and CME events are rare.

In recent years, stellar CMEs have attracted much attention for their significant impacts on mass and angular momentum loss in the course of stellar evolution \citep[e.g.,][]{Aarnio2012, Osten2015} and the potential habitability of the surrounding exoplanets \citep[e.g.,][]{Airapetian2016, Cherenkov2017, Hazra2022, Cohen2022}. Because direct imaging of stellar CMEs ---as performed on the Sun--- is impossible with current instrumentation, their detection is mainly based on other observational signatures known from studies of solar CMEs. Detailed reviews of the different methodologies used to attempt the detection of stellar CMEs can be found in  \citet{Leitzinger2022} and \citet{Tian2023}. In the optical wavelength regions, stellar CME candidates can be detected using the features of Doppler-shifted emission and absorption in the wings of sensitive chromospheric lines, especially the Balmer lines \citep[e.g.,][]{Houdebine1990, Vida2019, Leitzinger2020, Koller2021, Namekata2022, Lu2022, Wu2022, Inoue2023}. If the CME lies directly on the line of sight to the observer, it may produce an absorption signature \citep[see][]{Namekata2022}. In addition to obtaining spectroscopic observations of individual stars, research has also been conducted to detect possible stellar CMEs by investigating a large number of stars based on archival data \citep{Fuhrmeister2018, Vida2019, Leitzinger2020, Koller2021, Lu2022}.

It is widely acknowledged that the possible CME signatures identified in the optical spectral line profiles mostly correspond to eruptive prominences, which are believed to form the CME cores in the frequently observed three-part structure of solar CMEs \citep{Forbes2000}. With the exception of prominence eruptions and CMEs, the Doppler-shifted emission and absorption signatures observed in the chromospheric line profiles can also be attributed to other forms of plasma motion during the flares, including chromospheric evaporation \citep[e.g.,][]{Li2011, Tei2018, Honda2018}, chromospheric condensation \citep[e.g.,][]{Ichimoto1984, Tei2018, Vida2019}, and coronal rain \citep[e.g.,][]{Antolin2012, Lacatus2017, Li2021}.

In this study, we present the results of time-resolved high-resolution spectroscopic observations of one of the most active RS~CVn-type stars, II~Pegasi (II~Peg). It is notable that the H${\alpha}$ line profile shows flare-related asymmetric behavior during our observation, which implies the existence of moving plasmas in the outer atmosphere of II~Peg during the flare. This observation provides valuable insights that will contribute to the future detection and identification of stellar prominence eruptions and CMEs.

II~Peg (=~HD 224085) is a single-lined spectroscopic binary system, consisting of a K2~IV primary star and an unseen secondary companion with a period of about 6.7 days \citep{Berdyugina1998}. More detailed information regarding the physical parameters of II~Peg can be found in our previous paper \citep{Cao2024}. The K2~IV primary star shows remarkable photometric starspot activity \citep[e.g.,][]{Vogt1981, Berdyugina1998b, Rodono2000, Lindborg2013}, intense $\mbox{Ca~{\sc ii}}$~H\&K and H${\alpha}$ line emission \citep[e.g.,][]{Vogt1981, Huenemoerder1987, Montes1997, Berdyugina1999, Frasca2008}, and strong UV and X-ray radiation \citep[e.g.,][]{Byrne1989, Ercolano2008}. Moreover, II~Peg is a high-rate flaring star, showing flares several times in the optical, UV, and X-ray wavelength bands \citep[e.g.,][]{Rodono1987, Doyle1991, Berdyugina1999, Frasca2008, Ercolano2008, Siwak2010, Tsuboi2016, Cao2024}. Therefore, based on the close association between solar CMEs and flares,  II Peg is expected to produce frequent CME events. Recently, we detected a highly redshifted emission signature in the high-resolution H${\alpha}$ spectra during a stronger flare on II~Peg \citep{Cao2024}. The redshifted emission signature can be explained by a possible CME with backward propagation due to its fast Doppler-shift velocity or chromospheric condensation.

In Sect.~\ref{sec2}, we provide details of our high-resolution spectroscopic observation and data reduction. Our spectral analysis is described in Sect.~\ref{sec3}, especially for the H${\alpha}$ line. In Sect.~\ref{sec4}, we provide a more detailed discussion of a particular optical flare event and the asymmetric H${\alpha}$ line profile. Finally, we conclude and summarize our findings in Sect.~\ref{sec5}.
\begin{figure*}
\centering
\includegraphics[width=18cm,height=13cm]{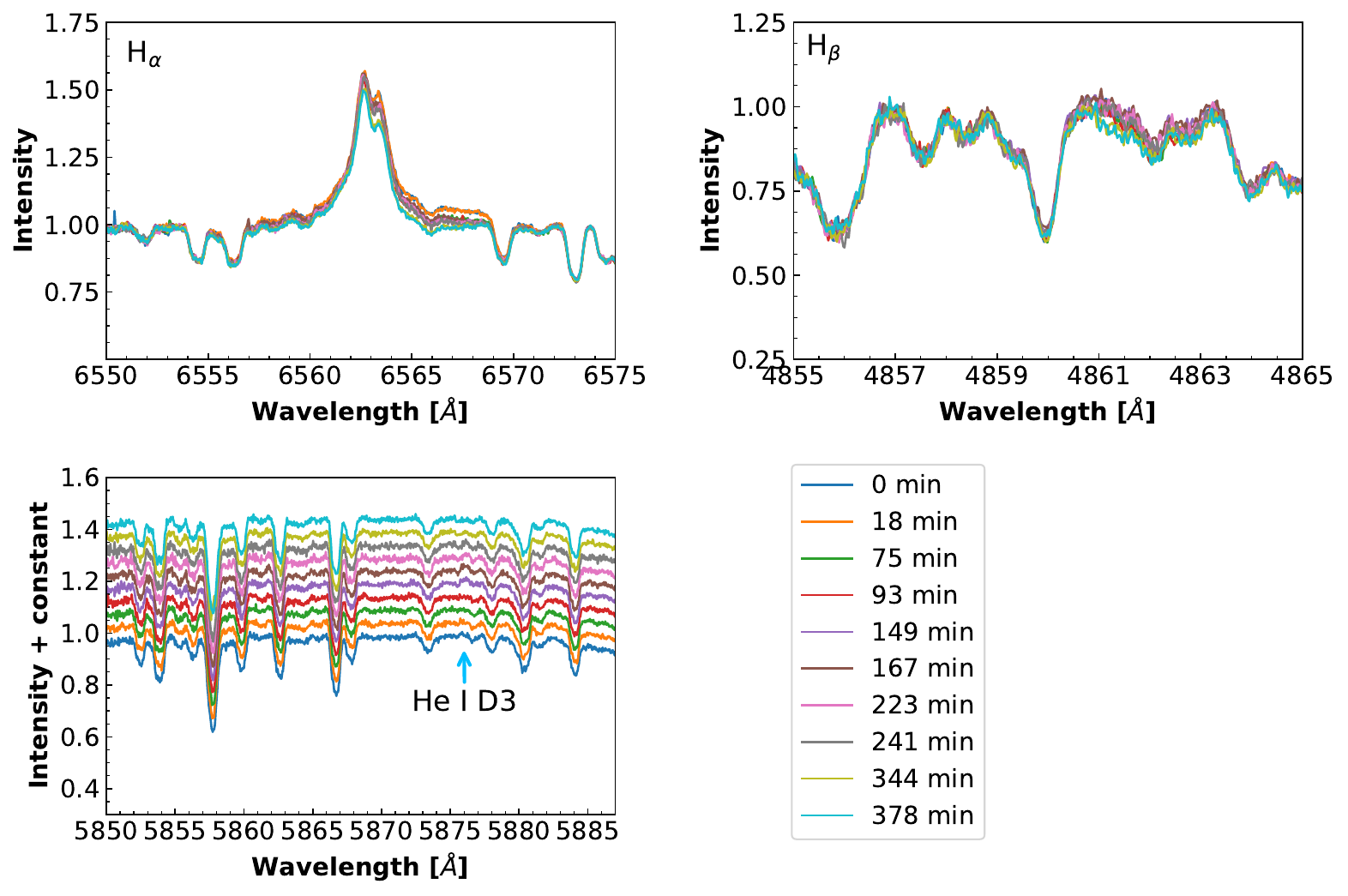}
\caption{Time variation of the H${\alpha}$,  H${\beta}$, and $\mbox{He~{\sc i}}$ D$_{3}$ line profiles of II~Peg observed on 2015~Oct~30. The $\mbox{He~{\sc i}}$ D$_{3}$ line profiles are shifted on the y-axis for illustrative purposes.}
\label{Fig1}
\end{figure*}
\section{Spectroscopic observation and data reduction}\label{sec2}
Spectra of II~Peg analyzed in the present work were obtained during the night of 2015~October~30. The observation was performed with the fiber-fed high-resolution spectrograph (HRS, \citealt{Fan2016}) installed on the 2.16m telescope at the Xinglong station of the National Astronomical Observatories, Chinese Academic of Sciences, China. HRS produces spectra with a resolving power of R~=~$\lambda$/$\Delta\lambda$~$\simeq$~48000 in a wavelength range of 3900--10000~\AA, using a $4096 \times 4096$ pixel CCD detector to record spectra. 

During our observation, a total of ten spectra of II~Peg were obtained over a six-hour period. The first eight spectra had an exposure time of 900~s, while the last two had an exposure time of 1800~s. The signal-to-noise ratio (S/N) at the continuum near the H${\alpha}$ line region was around 200. To clearly indicate the time variation of our observation, we set the observation time (the mid-exposure of observation) of the first spectrum to zero and then derived the observation time of the other spectra.

The data reduction was performed with the IRAF\footnote{IRAF is distributed by the National Optical Astronomy Observatories, which is operated by the Association of Universities for Research in Astronomy (AURA), Inc., under cooperative agreement with the National Science Foundation.} package, following the standard procedures \citep[see the prescription in][]{Cao2024}. Moreover, for our observation, there were heavy telluric water vapor lines in the activity line regions of interest, especially in the H${\alpha}$ and $\mbox{He~{\sc i}}$ D$_{3}$ line regions. We eliminated these using the spectrum of a brighter and rapidly rotating early-type star HR~8858 (spectral type: B5~V, $vsini$ = 316~km~s$^{-1}$) as a telluric line template, with an interactive procedure in the IRAF package.
\begin{figure*}
\sidecaption
\includegraphics[width=10.75cm,height=12cm]{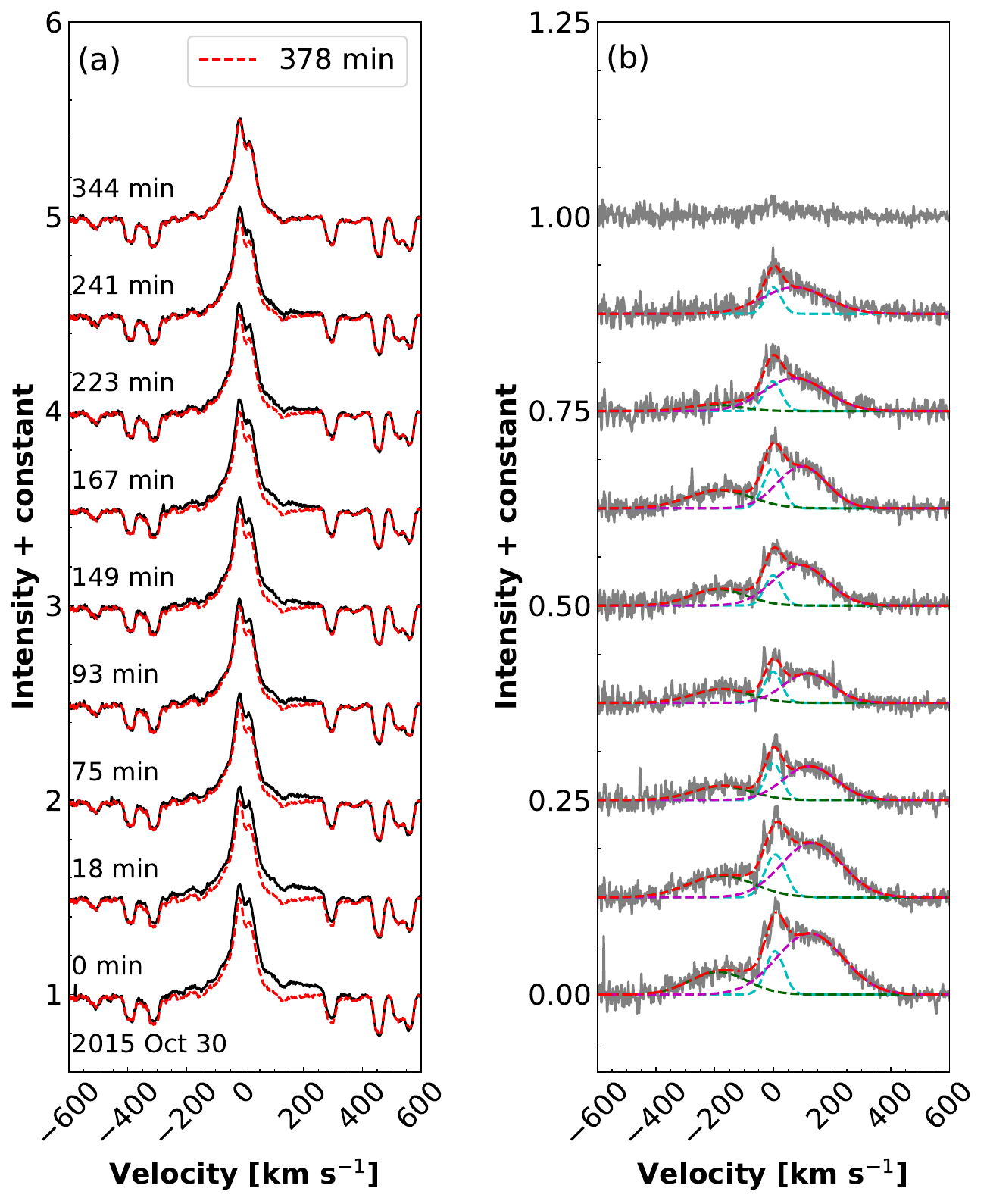}
\caption{Time variation of the H${\alpha}$ spectrum. (a) The first nine H${\alpha}$ spectra (black solid lines) are compared with the last one. The red dashed line represents the last spectrum observed at phase~0.809. (b) The H${\alpha}$ residuals between the first nine spectra and the last spectrum are shown and fitted using Gaussian fitting (dashed lines).}
\label{Fig2}
\end{figure*}
\begin{figure}
\centering
\includegraphics[width=9cm,height=12cm]{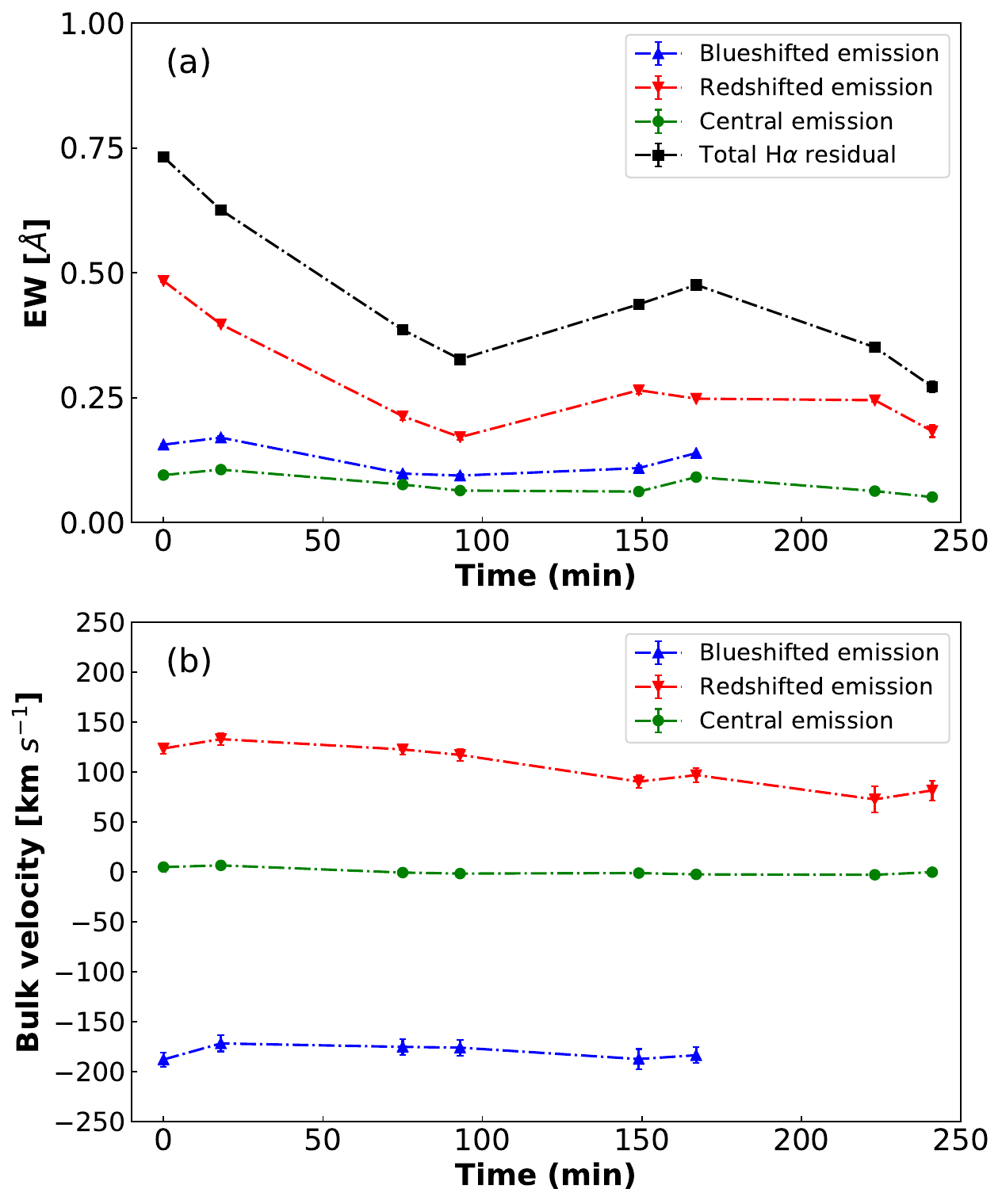}
\caption{Time evolution of the measurements for the total H${\alpha}$ residual and each fitted component during our observation. (a) EWs of the total H${\alpha}$ residual, and the blueshifted, central, and redshifted emission components. (b) The bulk velocities of the blueshifted, central, and redshifted emission components.}
\label{Fig3}
\end{figure}
\section{Spectral analysis}\label{sec3}
Figure~\ref{Fig1} shows the H${\alpha}$,  H${\beta}$, and $\mbox{He~{\sc i}}$ D$_{3}$ line profiles obtained during our observation, in which the radial velocity of all spectra are corrected so that they are aligned. From the figure, we can see that the H${\alpha}$ line shows an emission feature above the continuum level and has a significantly variable profile. In particular, the red wing gradually decreased from a strong enhancement to being almost flat. Moreover, the H${\beta}$ line profile is characterized by a deep absorption feature, but significant changes also found in its red wing, which is consistent with the behavior of the H${\alpha}$ line profile. Finally, it can be seen that the $\mbox{He~{\sc i}}$~D$_{3}$ line shows a weak emission feature in the first two spectra, and then the emission disappears in the next two spectra. From the fifth to the eighth spectrum, it is noteworthy that  a stronger emission feature reappears  in the $\mbox{He~{\sc i}}$~D$_{3}$ line, and then  disappears in the last two spectra.

To analyze the variation of the H${\alpha}$ line profile shown in Fig.~\ref{Fig2}~(a), we subtracted the last spectrum from the first nine. This is reasonable because the red wing of the last H${\alpha}$ profile had become flat, and there was no emission feature in the corresponding $\mbox{He~{\sc i}}$~D$_{3}$ line region, which means that there is no special activity at that time. Therefore, the last H${\alpha}$ profile can be used as a quiescence reference of the flare. We then obtained the quiescence reference-subtracted H${\alpha}$ line profiles (hereafter referred to as H${\alpha}$ residuals) shown in Fig.~\ref{Fig2}~(b). With the H${\alpha}$ residual, we model the line profile using a Gaussian fitting method. It can be seen that the profile can be well fitted by a weaker blueshifted broad emission component, a nearly centred narrow emission component, and a stronger redshifted broad emission component. From the figure, it is obvious that the intensity of the blueshifted emission component is weaker than the redshifted emission component, and that the duration of the blueshifted component is shorter than the redshifted one. Due to low S/N, however, we were not able to carry out any further analysis of the H${\beta}$ line.

We measured the equivalent widths (EWs) of the total H${\alpha}$ residual, and the blueshifted, central, and redshifted emission components, as well as their bulk velocities. Furthermore, we plot the time evolution of these measured values in Fig.~\ref{Fig3}.
\section{Results and discussion}\label{sec4}
\subsection{Optical flare}
Typically, solar and stellar flares can be observed across the entire electromagnetic spectrum from shorter X-rays to longer radio wavelengths \citep{Schrijver2000}. In the optical spectral range, the $\mbox{He~{\sc i}}$~D$_{3}$ line is an important and commonly used indicator to capture the flare activity events in solar and stellar chromospheres. Due to its very high excitation potential, the $\mbox{He~{\sc i}}$~D$_{3}$ line shows emission features above the continuum when an optical flare happens \citep[see][]{Zirin1988, Gu2002, Cao2019, Cao2023, Cao2024}. Moreover, the other chromospheric sensitivity lines, especially the H${\alpha}$ line, also show stronger emission and variable line profiles during the flares. Therefore, the appearing and disappearing features of the $\mbox{He~{\sc i}}$ D$_{3}$ line emission and the variation of the H${\alpha}$ line profile might indicate an evolved flare activity event during our observations. Due to the limited nature of our observations, we did not capture the entire flare evolution from the initial outburst to the end. From Fig.~\ref{Fig3}, it can be seen that the EW of the H${\alpha}$ residual shows a decreasing trend from the starting observation to 93 minutes, which is consistent with the disappearance of the $\mbox{He~{\sc i}}$~D$_{3}$ emission feature. Therefore, it is believed that we captured part of the gradual decay phase of an optical flare during our observation. 

Moreover, from Fig.~\ref{Fig3}, it is noteworthy that the EW of the H${\alpha}$ residual gradually increases between 93 and 167 minutes, and then decreases after 167 minutes, which implies the occurrence of a secondary peak during the flare decay phase. We were able to confirm the secondary peak by capturing the reappearance of the $\mbox{He~{\sc i}}$~D$_{3}$ line emission feature at the corresponding phases (see Fig.~\ref{Fig1} and the description in Sect.~\ref{sec3}).

We computed the stellar continuum flux $F_{H_{\alpha}}$~(in erg cm$^{-2}$ s$^{-1}$ \AA$^{-1}$) near the H${\alpha}$ line region as a function of the color index $B-V$ ($\sim$~1.031 for II~Peg; \citealt{Messina2008}) based on the empirical relationship: 
\begin{eqnarray}
\log{F_{H{\alpha}}}=[7.538-1.081(B-V)]\pm{0.33}\nonumber \\
0.0~\leq~B-V~\leq~1.4
\end{eqnarray} 
from \citet{Hall1996}, and then converted the EW of the H${\alpha}$ residual into an absolute surface flux $F_{S}$~(in erg~cm$^{-2}$~s$^{-1}$). Based on the derived flux $F_{S}$, we were able to calculate the luminosity $L_{H_{\alpha}}$~(in erg~s$^{-1}$) using the stellar radius $R_{\star} $~=~3.4$R_{\sun}$ \citep{Berdyugina1998} of the K2~IV primary star of II~Peg. Thus, we found the luminosity $L_{H_{\alpha}}$ of the flare to vary from $1.36 \times 10^{30}$~erg~s$^{-1}$ to $0.51 \times 10^{30}$~erg~s$^{-1}$ during our observation. Finally, integrating the luminosity of each flaring spectrum over the exposure time and then adding them together, the energy released in the H${\alpha}$ line is estimated to be approximately $6.03~\times~10^{33}$~erg, which is the lower limit of the flare energy. Moreover, according to the empirical relationship between the H${\alpha}$ (E$_{H{\alpha}}$) and bolometric white-light (E$_{WL, bol}$) flare energy from \citet{Namekata2024}:
\begin{eqnarray}
\frac{E_{WL, bol}}{10^{33}~erg}~=~10^{1.68\pm0.15}(\frac{E_{H{\alpha}}}{10^{33}~erg})^{1.04\pm0.07},
\end{eqnarray}
the bolometric white-light flare radiation energy, E$_{WL, bol}$, is calculated to be approximately $3.10~\times~10^{35}$~erg, which is comparable to the energy of stellar superflares, which release energies of more than 10$^{33}$~erg.

\subsection{Interpretation of the H${\alpha}$ asymmetry}
During our observation, the H${\alpha}$ residual shows asymmetry, including a blueshifted emission component and a redshifted emission one, which suggests the existence of moving plasmas in the atmosphere during the flare on II~Peg. Moreover, as shown in Fig.~\ref{Fig3}~(a), an increasing trend is seen in the intensities of the blueshifted, central, and redshifted emission components during the secondary peak; this trend is particularly evident for the redshifted emission component. The observed increase in the central emission component may indicates a secondary flare occurring during the decay phase of the main flare. Additionally, the observed increase in the blueshifted and redshifted emission components may be indicative of a notable increase in the scale of the moving plasmas.

\subsubsection{Blueshifted emission component}
Based on the current knowledge of solar flares, the blueshifted emission component could be ascribed to upward plasma flow above the chromospheric evaporation flows or an erupting prominence moving toward the observer \citep[e.g.,][]{Canfield1990, Li2011, Tei2018, Honda2018, Otsu2022}. 

In the case examined here, the blueshifted emission component has a bulk velocity of nearly -180~km~s$^{-1}$ (see Fig.~\ref{Fig3}) and extends its velocity to more than -350~km~s$^{-1}$ on the blue side. The higher Doppler velocity is likely indicative of a prominence eruption, and part of the prominence material may have escaped from the star, developing into a CME, given the maximum velocity of the blueshifted emission component, which exceeds the escape velocity ($\sim$~306~km~s$^{-1}$, see \citealt{Cao2024}) at the surface of the K2~IV primary star of II~Peg. Moreover, it is important to note that the velocity mentioned is only the projection component along the line of sight. Accordingly, if there were a larger angle between the eruption direction and the line of sight, the true eruption velocity would be significantly greater, thereby increasing the probability that the above interpretation is accurate.

Chromospheric evaporation usually occurs in the early phase of a solar flare and is caused by upward moving material driven by intense chromospheric heating, with typical velocities in the range of tens of kilometers per second \citep[e.g.,][]{Kennedy2015, Fuhrmeister2018, Tei2018}. For our situation, the velocity of the blueshifted emission component is greater than the typical upward velocity of solar chromospheric evaporation. Because the energy scale of the flare on II~Peg greatly exceeds that of typical solar flares, the magnitude of the blueshifted phenomena may be very different between solar flares and flares on stars. Therefore, it is also possible that the blueshifted emission component is caused by chromospheric evaporation. Moreover, the increasing intensity of the blueshifted emission component during the secondary peak may suggest that chromospheric evaporation becomes prominent, which is likely due to the additional heating of the chromosphere caused by the secondary flare.

\subsubsection{Redshifted emission component}
In contrast to the blueshifted emission component, the origin of the redshifted component is more complex. During our observation, the redshifted component has a bulk velocity of 130 to 70~km~s$^{-1}$ and extends its velocity to nearly 400~km~s$^{-1}$ on the red side (see Fig.~\ref{Fig3}~(b) and Fig.~\ref{Fig2}~(b)). According to our current knowledge of solar flares and the available literature for flares on other stars \citep[e.g.,][]{Fuhrmeister2018, Koller2021, Wu2022, Cao2024}, there are several possible interpretations of the redshifted emission component, including flare-driven coronal rain, chromospheric condensation, a backward-directed prominence eruption, or falling material in a prominence eruption.

Coronal rain is a frequently observed phenomenon in the solar atmosphere. Flare-driven coronal rain usually forms as a result of the radiative cooling process of hot flaring plasma in post-flare loops and then falls down along post-flare loops to the surface under the effect of gravity. Thus, coronal rain often becomes more pronounced during the decay phase of flares and plays a fundamental role in the mass cycle between the corona and chromosphere \citep[e.g.,][]{Li2021, Kohutova2019}. The typical velocities of coronal rain range from 30 to 200~km~s$^{-1}$, with an average value of about 60--70~km~s$^{-1}$ \citep[e.g.,][]{Antolin2012, Lacatus2017}. The redshifted emission component seen in the observations presented here has the observational characteristics of coronal rain and the post-flare loops might become prominent during the secondary peak.

Chromospheric condensation is usually observed during the impulsive phases of solar flares and is caused by downward-moving material driven by the nonthermal electron beams of the flares. The typical downward mass velocity is approximately tens of kilometers per second, and sometimes even reaches or exceeds 100~km~s$^{-1}$ \citep{Ichimoto1984, Tei2018}. As opposed to our interpretation of the blueshifted emission component as chromospheric evaporation, the redshifted emission component could be attributed to chromospheric condensation, and the likelihood of this phenomenon is particularly elevated during the secondary peak, which is due to the occurrence of a secondary flare.

A backward-directed prominence eruption could also result in the redshifted emission component and a detectable backward-directed prominence eruption requires special spatial location \citep[e.g.,][]{Moschou2019, Koller2021, Wu2022}. In our previous paper \citep{Cao2024}, we reported a possible backward-directed prominence eruption event, showing the redshifted emission component in the high-resolution spectra with a large bulk velocity of about 429~km~s$^{-1}$. The large velocity greatly exceeds the escape velocity of the K2~IV star of II~Peg, which indicates that the prominence eruption can develop into a CME. In the present case, the redshifted emission component could also be attributed to a backward-directed prominence eruption near the stellar limb. On the one hand, similar to the blueshifted emission component, the velocity is only the component along the line of sight. When the prominence eruption occurs close to the stellar limb, the eruption velocity would be much greater than the measured Doppler velocity. Furthermore, the maximum velocity of the reshifted emission component extends beyond the escape velocity of II~Peg. Therefore, it is possible that this prominence eruption could also develop into a CME.

Strong magnetic field on active stars may prevent prominence material from leaving the stellar surface \citep{Drake2016, Alvarado2018}, thus falling material in a prominence eruption would also result in a redshifted emission component. \citet{Wu2022} depicted a similar scenario where the erupted prominence material may be confined by an overlying magnetic field, subsequently exhibiting downward material falling onto the stellar disk. This scenario could potentially account for the redshifted emission component in our observation. Moreover, the falling material would also result in flare heating and \textcolor[rgb]{0,0,0}{this would explain the occurrence of} a secondary flare during the flare decay. For instance, \citet{Li2017} reported a variety of flare brightenings associated with falling prominence material in the extreme-ultraviolet (EUV) passband on the Sun, including the two ribbon-like brightenings caused by the prominence material hitting the chromosphere when it falls back to the solar surface. These authors also reported brightenings resulting from the magnetic reconnection between the prominence magnetic structure and the surrounding magnetic field, and a thread-like brightening on top of the erupting magnetic field caused by an energy imbalance from a fast drop of radiative cooling due to plasma rarefaction. As shown in Fig.~\ref{Fig3} (b), the velocity of the redshifted emission component we see appears to gradually slow down, which is probably because falling prominence material was influenced by the surrounding magnetic field and therefore slowed down by magnetic drag force \citep{Li2017}. Thus, the magnetic reconnection between the prominence magnetic structure and the surrounding magnetic field may be the main mechanism triggering the occurrence of the secondary flare during our observation. Although the flare brightenings described by \citet{Li2017} were observed in the EUV passband, it is possible that similar flare brightenings could also be found in the H${\alpha}$ line, because flares usually produce radiation across the entire electromagnetic spectrum. Finally, as stated by \citet{Wu2022}, prior to the detection of the red asymmetry caused by falling material in a prominence eruption, there should be a short-lived blueshift ascribed to the ejecting prominence material at the flare onset. In our case, the entire life cycle of the flare was not captured by our observations. Therefore, we are not sure whether or not there was a blueshift at the flare onset. For our situation, if the blueshifted emission component is caused by the prominence eruption, we can assume it might be the continuation of the one that occurred at the flare onset and led to prominence material falling during our observation. However, the velocity of the blueshifted emission component is nearly stable and shows no decreasing trend (see Fig.~\ref{Fig3} (b)), probably because the prominence eruption is not a brief process and comprises an array of materials with different velocities.

Stellar flares are highly dramatic explosions that can be accompanied by prominence eruptions, CMEs, and other forms of plasma motion. As a result, the observed H${\alpha}$ features during the flare may represent a combination of the flare itself, a prominence eruption, a CME, and other plasma motions. Taking into consideration the evolutionary behavior of the redshifted emission component in the H${\alpha}$ residual of II~Peg, it is possible that this phenomenon might also be caused by a combination of some of the aforementioned phenomena. For instance, in the scenarios of backward-directed prominence eruptions and falling material in prominence eruptions, the increasing scale of the redshifted emission component may be attributed to other forms of plasma motion, such as coronal rain or chromospheric condensation.

\subsection{Masses of the moving plasmas}
Regarding the observations studied here, we were able to estimate the masses of the moving plasma material generating the blueshifted and  redshifted emission components. By following \citet{Houdebine1990} and \citet{Koller2021}, the minimum mass of the moving plasma can be expressed as:
\begin{eqnarray}
M~\geqslant~\frac{4{\pi}d^{2}f_{line}m_{H}}{A_{ji}h\nu_{ji}P_{esc}}\frac{N_{tot}}{N_{j}}
,\end{eqnarray}
where $d$ is the distance from the star, $f_{line}$ is the corresponding line flux from the Doppler-shifted emission feature, $m_{H}$ is the mass of a hydrogen atom, $N_{tot}/N_{j}$ is the ratio between the number density of hydrogen atoms and the number density of hydrogen atoms at excited level $j$, $h$ is the Planck constant, $\nu_{ji}$ and $A_{ji}$ are the frequency and Einstein coefficient for a spontaneous decay from level $j$ to $i$, and $P_{esc}$ is the escape probability. 

We adopt the same parameter settings and calculation strategy as \citet{Koller2021}. Moreover, according to the strategy of \citet{Wu2022}, we use the luminosity $L_{line}$ from the Doppler-shifted emission instead of $4{\pi}d^{2}f_{line}$ as we have no absolute flux calibration during our observation. Based on the EWs at the beginning of our observation, the luminosities of the blueshifted and redshifted emission components are calculated to be $2.9 \times 10^{29}$~erg~s$^{-1}$ and $9.0 \times 10^{29}$~erg~s$^{-1}$, respectively. Therefore, we derive lower limits on the masses of the moving plasma materials of $0.56 \times 10^{20}$~g and $1.74 \times 10^{20}$~g, respectively. These masses are indeed in agreement with literature values of eruptive prominences seen on stars \citep[e.g.,][]{Namekata2022, Inoue2023}.

\section{Summary and conclusions}\label{sec5}
Part of the gradual decay phase of an optical flare was detected during our observation, and the low limit of the flare energy released in the H${\alpha}$ line is estimated to be $6.03~\times~10^{33}$~erg. We converted this H${\alpha}$ energy into bolometric white-light flare energy, finding a value of $3.10~\times~10^{35}$~erg , which suggests that the flare could be classified as a superflare. Moreover, a secondary peak was observed during the flare decay. After removing a quiescence reference spectrum, the H${\alpha}$ residual shows an asymmetrical behavior, including a blueshifted and a redshifted emission component, which implies the existence of moving plasma in the atmosphere of II~Peg.

The blueshifted emission component has a bulk velocity of about -180~km~s$^{-1}$ and extends its velocity to more than -350~km~s$^{-1}$. This component can be attributed to a possible prominence eruption due to its high velocity. Moreover, an alternative interpretation of the blueshifted emission component is chromospheric evaporation. The redshifted emission component is indicative of plasma moving away from the observer with a bulk velocity ranging from 130 to 70~km~s$^{-1}$, and extends its velocity to nearly 400~km~s$^{-1}$. Based on our knowledge of solar flares and the currently literature regarding flares on other stars, there could be several explanations for the plasma moving away from us, or indeed a combination of  some of them: a flare-driven coronal rain process, a chromospheric condensation process, a backward-directed prominence eruption happening near the stellar limb, or falling material in a prominence eruption. The minimum masses of the moving plasmas resulting in the blueshifted and redshifted emission components are estimated to be $0.56 \times 10^{20}$~g and $1.74 \times 10^{20}$~g, respectively.

\begin{acknowledgements}
The authors express their gratitude to the staff of the Xinglong 2.16 m telescope for their assistance and support during the observation. This work is partially supported by the Open Project Program of the Key Laboratory of Optical Astronomy, National Astronomical Observatories, Chinese Academy of Sciences. The present study is also financially supported by the National Natural Science Foundation of China (NSFC) under grant Nos. 10373023, 10773027, 11333006, and U1531121, the Yunnan Fundamental Research Projects (grant Nos.~202201AT070186 and 202305AS350009), the Yunnan Revitalization Talent Support Program (Young Talent Project), and International Centre of Supernovae, Yunnan Key Laboratory (No.~202302AN360001). The authors also acknowledge the science research grant from the China Manned Space Project.
\end{acknowledgements}

%
\bibliographystyle{aa}
\bibliography{aa} 

@ARTICLE{Montes1997,
       author = {{Montes}, D. and {Fernandez-Figueroa}, M.~J. and {de Castro}, E. and {Sanz-Forcada}, J.},
        title = "{Multiwavelength optical observations of chromospherically active binary systems. I. Simultaneous H{\ensuremath{\alpha}}, NA I D$_{1}$, D$_{2}$, and He I D$_{3}$ observations}",
      journal = {\aaps},
         year = 1997,
        month = oct,
       volume = {125},
        pages = {263-287},
          doi = {10.1051/aas:1997374},
archivePrefix = {arXiv},
       eprint = {astro-ph/9702127},
 primaryClass = {astro-ph},
       adsurl = {https://ui.adsabs.harvard.edu/abs/1997A&AS..125..263M}
}

@article{Berdyugina1999,
       author = {{Berdyugina}, S.~V. and {Ilyin}, I. and {Tuominen}, I.},
        title = "{The active RS Canum Venaticorum binary II Pegasi. III. Chromospheric emission and flares in 1994-1996}",
      journal = {\aap},
         year = 1999,
        month = sep,
       volume = {349},
        pages = {863-872},
       adsurl = {https://ui.adsabs.harvard.edu/abs/1999A&A...349..863B}
}

@article{Siwak2010,
       author = {{Siwak}, Michal and {Rucinski}, Slavek M. and {Matthews}, Jaymie M. and {Kuschnig}, Rainer and {Guenther}, David B. and {Moffat}, Anthony F.~J. and {Sasselov}, Dimitar and {Weiss}, Werner W.},
        title = "{Analysis of the MOST light curve of the heavily spotted K2IV component of the single-line spectroscopic binary II Pegasi}",
      journal = {\mnras},
         year = 2010,
        month = oct,
       volume = {408},
       number = {1},
        pages = {314-321},
          doi = {10.1111/j.1365-2966.2010.17109.x},
archivePrefix = {arXiv},
       eprint = {1009.1171},
 primaryClass = {astro-ph.SR},
       adsurl = {https://ui.adsabs.harvard.edu/abs/2010MNRAS.408..314S}
}

@article{Tsuboi2016,
       author = {{Tsuboi}, Yohko and {Yamazaki}, Kyohei and {Sugawara}, Yasuharu and {Kawagoe}, Atsushi and {Kaneto}, Soichiro and {Iizuka}, Ryo and {Matsumura}, Takanori and {Nakahira}, Satoshi and {Higa}, Masaya and {Matsuoka}, Masaru and {Sugizaki}, Mutsumi and {Ueda}, Yoshihiro and {Kawai}, Nobuyuki and {Morii}, Mikio and {Serino}, Motoko and {Mihara}, Tatehiro and {Tomida}, Hiroshi and {Ueno}, Shiro and {Negoro}, Hitoshi and {Daikyuji}, Arata and {Ebisawa}, Ken and {Eguchi}, Satoshi and {Hiroi}, Kazuo and {Ishikawa}, Masaki and {Isobe}, Naoki and {Kawasaki}, Kazuyoshi and {Kimura}, Masashi and {Kitayama}, Hiroki and {Kohama}, Mitsuhiro and {Kotani}, Taro and {Nakagawa}, Yujin E. and {Nakajima}, Motoki and {Ozawa}, Hiroshi and {Shidatsu}, Megumi and {Sootome}, Tetsuya and {Sugimori}, Kousuke and {Suwa}, Fumitoshi and {Tsunemi}, Hiroshi and {Usui}, Ryuichi and {Yamamoto}, Takayuki and {Yamaoka}, Kazutaka and {Yoshida}, Atsumasa},
        title = "{Large X-ray flares on stars detected with MAXI/GSC: A universal correlation between the duration of a flare and its X-ray luminosity}",
      journal = {\pasj},
         year = 2016,
        month = oct,
       volume = {68},
       number = {5},
          eid = {90},
        pages = {90},
          doi = {10.1093/pasj/psw081},
archivePrefix = {arXiv},
       eprint = {1609.01925},
 primaryClass = {astro-ph.HE},
       adsurl = {https://ui.adsabs.harvard.edu/abs/2016PASJ...68...90T}
}

@ARTICLE{Koller2021,
       author = {{Koller}, Florian and {Leitzinger}, Martin and {Temmer}, Manuela and {Odert}, Petra and {Beck}, Paul G. and {Veronig}, Astrid},
        title = "{Search for flares and associated CMEs on late-type main-sequence stars in optical SDSS spectra}",
      journal = {\aap},
         year = 2021,
        month = feb,
       volume = {646},
          eid = {A34},
        pages = {A34},
          doi = {10.1051/0004-6361/202039003},
archivePrefix = {arXiv},
       eprint = {2012.00786},
 primaryClass = {astro-ph.SR},
       adsurl = {https://ui.adsabs.harvard.edu/abs/2021A&A...646A..34K}
}

@ARTICLE{Houdebine1990,
       author = {{Houdebine}, E.~R. and {Foing}, B.~H. and {Rodono}, M.},
        title = "{Dynamics of flares on late-type dMe stars. I. Flare mass ejections and stellar evolution.}",
      journal = {\aap},
         year = 1990,
        month = nov,
       volume = {238},
        pages = {249},
       adsurl = {https://ui.adsabs.harvard.edu/abs/1990A&A...238..249H}
}

@ARTICLE{Hall1996,
       author = {{Hall}, Jeffrey C.},
        title = "{On the Determination of Empirical Stellar Flux Scales}",
      journal = {\pasp},
         year = 1996,
        month = apr,
       volume = {108},
        pages = {313},
          doi = {10.1086/133724},
       adsurl = {https://ui.adsabs.harvard.edu/abs/1996PASP..108..313H}
}

@BOOK{Zirin1988,
       author = {{Zirin}, Harold},
        title = "{Astrophysics of the sun}",
	publisher = {Cambridge Univ. Press, Cambridge},
         year = 1988,
       adsurl = {https://ui.adsabs.harvard.edu/abs/1988assu.book.....Z}
}

@ARTICLE{Messina2008,
       author = {{Messina}, S.},
        title = "{Long-term magnetic activity in close binary systems. I. Patterns of color variations}",
      journal = {\aap},
         year = 2008,
        month = mar,
       volume = {480},
       number = {2},
        pages = {495-508},
          doi = {10.1051/0004-6361:20078932},
archivePrefix = {arXiv},
       eprint = {0712.2117},
 primaryClass = {astro-ph},
       adsurl = {https://ui.adsabs.harvard.edu/abs/2008A&A...480..495M}
}

@ARTICLE{Berdyugina1998,
       author = {{Berdyugina}, S.~V. and {Jankov}, S. and {Ilyin}, I. and {Tuominen}, I. and {Fekel}, F.~C.},
        title = "{The active RS Canum Venaticorum binary II Pegasi. I. Stellar and orbital parameters}",
      journal = {\aap},
         year = 1998,
        month = jun,
       volume = {334},
        pages = {863-872},
       adsurl = {https://ui.adsabs.harvard.edu/abs/1998A&A...334..863B}
}

@ARTICLE{Cao2019,
       author = {{Cao}, Dongtao and {Gu}, Shenghong and {Ge}, Jian and {Wang}, Tinggui and {Zhou}, Jilin and {Chang}, Liang and {Wolter}, U. and {Mittag}, M. and {Schmitt}, J.~H.~M.~M. and {Perdelwitz}, V.},
        title = "{Prominence activation, optical flare, and post-flare loops on the RS Canum Venaticorum star SZ Piscium}",
      journal = {\mnras},
         year = 2019,
        month = jan,
       volume = {482},
       number = {1},
        pages = {988-998},
          doi = {10.1093/mnras/sty2768},
       adsurl = {https://ui.adsabs.harvard.edu/abs/2019MNRAS.482..988C}
}

@ARTICLE{Gu2002,
       author = {{Gu}, S. -H. and {Tan}, H. -S. and {Shan}, H. -G. and {Zhang}, F. -H.},
        title = "{Chromospheric activity on the RS CVn-type binary UX Arietis}",
      journal = {\aap},
         year = 2002,
        month = jun,
       volume = {388},
        pages = {889-898},
          doi = {10.1051/0004-6361:20020561},
       adsurl = {https://ui.adsabs.harvard.edu/abs/2002A&A...388..889G}
}

@ARTICLE{Frasca2008,
       author = {{Frasca}, A. and {Biazzo}, K. and {Ta{\c{s}}}, G. and {Evren}, S. and {Lanzafame}, A.~C.},
        title = "{Spots, plages, and flares on {\ensuremath{\lambda}} Andromedae and II Pegasi}",
      journal = {\aap},
         year = 2008,
        month = feb,
       volume = {479},
       number = {2},
        pages = {557-565},
          doi = {10.1051/0004-6361:20077915},
archivePrefix = {arXiv},
       eprint = {0711.3322},
 primaryClass = {astro-ph},
       adsurl = {https://ui.adsabs.harvard.edu/abs/2008A&A...479..557F}
}

@ARTICLE{Wu2022,
       author = {{Wu}, Yuchuan and {Chen}, Hechao and {Tian}, Hui and {Zhang}, Liyun and {Shi}, Jianrong and {He}, Han and {Lu}, Hongpeng and {Xu}, Yu and {Wang}, Haifeng},
        title = "{Broadening and Redward Asymmetry of H{\ensuremath{\alpha}} Line Profiles Observed by LAMOST during a Stellar Flare on an M-type Star}",
      journal = {\apj},
         year = 2022,
        month = apr,
       volume = {928},
       number = {2},
          eid = {180},
        pages = {180},
          doi = {10.3847/1538-4357/ac5897},
archivePrefix = {arXiv},
       eprint = {2203.02292},
 primaryClass = {astro-ph.SR},
       adsurl = {https://ui.adsabs.harvard.edu/abs/2022ApJ...928..180W}
}

@ARTICLE{Moschou2019,
       author = {{Moschou}, Sofia-Paraskevi and {Drake}, Jeremy J. and {Cohen}, Ofer and {Alvarado-G{\'o}mez}, Juli{\'a}n D. and {Garraffo}, Cecilia and {Fraschetti}, Federico},
        title = "{The Stellar CME-Flare Relation: What Do Historic Observations Reveal?}",
      journal = {\apj},
         year = 2019,
        month = jun,
       volume = {877},
       number = {2},
          eid = {105},
        pages = {105},
          doi = {10.3847/1538-4357/ab1b37},
archivePrefix = {arXiv},
       eprint = {1904.09598},
 primaryClass = {astro-ph.SR},
       adsurl = {https://ui.adsabs.harvard.edu/abs/2019ApJ...877..105M}
}

@ARTICLE{Lindborg2013,
       author = {{Lindborg}, M. and {Mantere}, M.~J. and {Olspert}, N. and {Pelt}, J. and {Hackman}, T. and {Henry}, G.~W. and {Jetsu}, L. and {Strassmeier}, K.~G.},
        title = "{Multiperiodicity, modulations and flip-flops in variable star light curves. II. Analysis of II Pegasus photometry during 1979-2010}",
      journal = {\aap},
         year = 2013,
        month = nov,
       volume = {559},
          eid = {A97},
        pages = {A97},
          doi = {10.1051/0004-6361/201321695},
archivePrefix = {arXiv},
       eprint = {1304.4144},
 primaryClass = {astro-ph.SR},
       adsurl = {https://ui.adsabs.harvard.edu/abs/2013A&A...559A..97L}
}

@ARTICLE{Byrne1989,
       author = {{Byrne}, P.~B. and {Panagi}, P. and {Doyle}, J.~G. and {Englebrecht}, C.~A. and {McMahan}, R. and {Marang}, F. and {Wegner}, G.},
        title = "{Rotational modulation and flares on RS Canum Venaticorum and BY Draconis stars. XII. Near-to-simultaneous high resolution UV and optical observations of II Pegasi during July 1984.}",
      journal = {\aap},
         year = 1989,
        month = apr,
       volume = {214},
        pages = {227-238},
       adsurl = {https://ui.adsabs.harvard.edu/abs/1989A&A...214..227B}
}

@ARTICLE{Rodono1987,
       author = {{Rodono}, M. and {Byrne}, P.~B. and {Neff}, J.~E. and {Linsky}, J.~L. and {Simon}, T. and {Butler}, C.~J. and {Catalano}, S. and {Cutispoto}, G. and {Doyle}, J.~G. and {Andrews}, A.~D. and {Gibson}, D.~M.},
        title = "{Rotational modulation and flares on RS CVn and BY DRA stars. III. IUEobserations of V711 Tau = (HR 1099), II Peg and AR Lac.}",
      journal = {\aap},
         year = 1987,
        month = apr,
       volume = {176},
        pages = {267-284},
       adsurl = {https://ui.adsabs.harvard.edu/abs/1987A&A...176..267R}
}

@ARTICLE{Doyle1991,
       author = {{Doyle}, J.~G. and {Kellett}, B.~J. and {Byrne}, P.~B. and {Avgoloupis}, S. and {Mavridis}, L.~N. and {Seiradakis}, J.~H. and {Bromage}, G.~E. and {Tsuru}, T. and {Makishima}, K. and {Makishima}, K. and {McHardy}, I.~M.},
        title = "{Simultaneous detection of a large flare in the X-ray and optical regions on the RS CVn-type star II Peg.}",
      journal = {\mnras},
         year = 1991,
        month = feb,
       volume = {248},
        pages = {503},
          doi = {10.1093/mnras/248.3.503},
       adsurl = {https://ui.adsabs.harvard.edu/abs/1991MNRAS.248..503D}
}

@ARTICLE{Huenemoerder1987,
       author = {{Huenemoerder}, David P. and {Ramsey}, Lawrence W.},
        title = "{CCD Echelle Observations of the Active RS CVn System II Pegasi}",
      journal = {\apj},
         year = 1987,
        month = aug,
       volume = {319},
        pages = {392},
          doi = {10.1086/165463},
       adsurl = {https://ui.adsabs.harvard.edu/abs/1987ApJ...319..392H}
}

@ARTICLE{Vogt1981,
       author = {{Vogt}, S.~S.},
        title = "{A spectroscopic photometric, and magnetic study of the starspot on IIPeg.}",
      journal = {\apj},
         year = 1981,
        month = aug,
       volume = {247},
        pages = {975-983},
          doi = {10.1086/159107},
       adsurl = {https://ui.adsabs.harvard.edu/abs/1981ApJ...247..975V}
}

@ARTICLE{Namekata2022,
       author = {{Namekata}, Kosuke and {Maehara}, Hiroyuki and {Honda}, Satoshi and {Notsu}, Yuta and {Okamoto}, Soshi and {Takahashi}, Jun and {Takayama}, Masaki and {Ohshima}, Tomohito and {Saito}, Tomoki and {Katoh}, Noriyuki and {Tozuka}, Miyako and {Murata}, Katsuhiro L. and {Ogawa}, Futa and {Niwano}, Masafumi and {Adachi}, Ryo and {Oeda}, Motoki and {Shiraishi}, Kazuki and {Isogai}, Keisuke and {Seki}, Daikichi and {Ishii}, Takako T. and {Ichimoto}, Kiyoshi and {Nogami}, Daisaku and {Shibata}, Kazunari},
        title = "{Probable detection of an eruptive filament from a superflare on a solar-type star}",
      journal = {Nature Astronomy},
         year = 2021,
        month = dec,
       volume = {6},
        pages = {241-248},
          doi = {10.1038/s41550-021-01532-8},
archivePrefix = {arXiv},
       eprint = {2112.04808},
 primaryClass = {astro-ph.SR},
       adsurl = {https://ui.adsabs.harvard.edu/abs/2022NatAs...6..241N}
}

@ARTICLE{Inoue2023,
       author = {{Inoue}, Shun and {Maehara}, Hiroyuki and {Notsu}, Yuta and {Namekata}, Kosuke and {Honda}, Satoshi and {Namizaki}, Keiichi and {Nogami}, Daisaku and {Shibata}, Kazunari},
        title = "{Detection of a High-velocity Prominence Eruption Leading to a CME Associated with a Superflare on the RS CVn-type Star V1355 Orionis}",
      journal = {\apj},
         year = 2023,
        month = may,
       volume = {948},
       number = {1},
          eid = {9},
        pages = {9},
          doi = {10.3847/1538-4357/acb7e8},
archivePrefix = {arXiv},
       eprint = {2301.13453},
 primaryClass = {astro-ph.SR},
       adsurl = {https://ui.adsabs.harvard.edu/abs/2023ApJ...948....9I}
}

@ARTICLE{Ercolano2008,
       author = {{Ercolano}, Barbara and {Drake}, Jeremy J. and {Reale}, Fabio and {Testa}, Paola and {Miller}, Jon M.},
        title = "{Fe K{\ensuremath{\alpha}} and Hydrodynamic Loop Model Diagnostics for a Large Flare on II Pegasi}",
      journal = {\apj},
         year = 2008,
        month = dec,
       volume = {688},
       number = {2},
        pages = {1315-1319},
          doi = {10.1086/591934},
archivePrefix = {arXiv},
       eprint = {0807.2093},
 primaryClass = {astro-ph},
       adsurl = {https://ui.adsabs.harvard.edu/abs/2008ApJ...688.1315E}
}

@ARTICLE{Schrijver2000,
       author = {{Schrijver}, C.~J. and {Zwaan}, C.},
        title = "{Solar and stellar magnetic activity.}",
      journal = {Cambridge Astrophysics Series},
         year = 2000,
        month = jan,
       volume = {34},
       adsurl = {https://ui.adsabs.harvard.edu/abs/2000CAS....34.....S}
}

@ARTICLE{Ichimoto1984,
       author = {{Ichimoto}, K. and {Kurokawa}, H.},
        title = "{H{\ensuremath{\alpha}} Red Asymmetry of Solar Flares}",
      journal = {\solphys},
         year = 1984,
        month = jun,
       volume = {93},
       number = {1},
        pages = {105-121},
          doi = {10.1007/BF00156656},
       adsurl = {https://ui.adsabs.harvard.edu/abs/1984SoPh...93..105I}
}

@ARTICLE{Cao2023,
       author = {{Cao}, Dongtao and {Gu}, Shenghong and {Wolter}, U. and {Mittag}, M. and {Schmitt}, J.~H.~M.~M. and {Gao}, Dongyang and {Hu}, Shaoming},
        title = "{Prominence detection and chromosphere feature on the prototype RS CVn of active binary systems}",
      journal = {\mnras},
         year = 2023,
        month = jun,
          doi = {10.1093/mnras/stad1700},
       adsurl = {https://ui.adsabs.harvard.edu/abs/2023MNRAS.tmp.1664C}
}

@INPROCEEDINGS{Drake2016,
       author = {{Drake}, Jeremy J. and {Cohen}, Ofer and {Garraffo}, Cecilia and {Kashyap}, V.},
        title = "{Stellar flares and the dark energy of CMEs}",
    booktitle = {Solar and Stellar Flares and their Effects on Planets},
         year = 2016,
       editor = {{Kosovichev}, A.~G. and {Hawley}, S.~L. and {Heinzel}, P.},
       volume = {320},
        month = jan,
        pages = {196-201},
          doi = {10.1017/S1743921316000260},
archivePrefix = {arXiv},
       eprint = {1610.05185},
 primaryClass = {astro-ph.SR},
       adsurl = {https://ui.adsabs.harvard.edu/abs/2016IAUS..320..196D}
}

@ARTICLE{Alvarado2018,
       author = {{Alvarado-G{\'o}mez}, Juli{\'a}n D. and {Drake}, Jeremy J. and {Cohen}, Ofer and {Moschou}, Sofia P. and {Garraffo}, Cecilia},
        title = "{Suppression of Coronal Mass Ejections in Active Stars by an Overlying Large-scale Magnetic Field: A Numerical Study}",
      journal = {\apj},
         year = 2018,
        month = aug,
       volume = {862},
       number = {2},
          eid = {93},
        pages = {93},
          doi = {10.3847/1538-4357/aacb7f},
archivePrefix = {arXiv},
       eprint = {1806.02828},
 primaryClass = {astro-ph.SR},
       adsurl = {https://ui.adsabs.harvard.edu/abs/2018ApJ...862...93A}
}

@ARTICLE{Fan2016,
       author = {{Fan}, Zhou and {Wang}, Huijuan and {Jiang}, Xiaojun and {Wu}, Hong and {Li}, Hongbin and {Huang}, Yang and {Xu}, Dawei and {Hu}, Zhongwen and {Zhu}, Yinan and {Wang}, Jianfeng and {Komossa}, Stefanie and {Zhang}, Xiaoming},
        title = "{The Xinglong 2.16-m Telescope: Current Instruments and Scientific Projects}",
      journal = {\pasp},
         year = 2016,
        month = nov,
       volume = {128},
       number = {969},
        pages = {115005},
          doi = {10.1088/1538-3873/128/969/115005},
archivePrefix = {arXiv},
       eprint = {1605.09166},
 primaryClass = {astro-ph.IM},
       adsurl = {https://ui.adsabs.harvard.edu/abs/2016PASP..128k5005F}
}

@ARTICLE{Antolin2012,
       author = {{Antolin}, Patrick and {Vissers}, Gregal and {Rouppe van der Voort}, Luc},
        title = "{On-Disk Coronal Rain}",
      journal = {\solphys},
         year = 2012,
        month = oct,
       volume = {280},
       number = {2},
        pages = {457-474},
          doi = {10.1007/s11207-012-9979-7},
       adsurl = {https://ui.adsabs.harvard.edu/abs/2012SoPh..280..457A}
}

@ARTICLE{Lacatus2017,
       author = {{Lacatus}, Daniela A. and {Judge}, Philip G. and {Donea}, Alina},
        title = "{An Explanation of Remarkable Emission-line Profiles in Post-flare Coronal Rain}",
      journal = {\apj},
         year = 2017,
        month = jun,
       volume = {842},
       number = {1},
          eid = {15},
        pages = {15},
          doi = {10.3847/1538-4357/aa725d},
       adsurl = {https://ui.adsabs.harvard.edu/abs/2017ApJ...842...15L}
}

@ARTICLE{Airapetian2016,
       author = {{Airapetian}, V.~S. and {Glocer}, A. and {Gronoff}, G. and {H{\'e}brard}, E. and {Danchi}, W.},
        title = "{Prebiotic chemistry and atmospheric warming of early Earth by an active young Sun}",
      journal = {Nature Geoscience},
         year = 2016,
        month = jun,
       volume = {9},
       number = {6},
        pages = {452-455},
          doi = {10.1038/ngeo2719},
       adsurl = {https://ui.adsabs.harvard.edu/abs/2016NatGe...9..452A}
}

@ARTICLE{Hazra2022,
       author = {{Hazra}, Gopal and {Vidotto}, Aline A. and {Carolan}, Stephen and {Villarreal D'Angelo}, Carolina and {Manchester}, Ward},
        title = "{The impact of coronal mass ejections and flares on the atmosphere of the hot Jupiter HD189733b}",
      journal = {\mnras},
         year = 2022,
        month = feb,
       volume = {509},
       number = {4},
        pages = {5858-5871},
          doi = {10.1093/mnras/stab3271},
archivePrefix = {arXiv},
       eprint = {2111.04531},
 primaryClass = {astro-ph.EP},
       adsurl = {https://ui.adsabs.harvard.edu/abs/2022MNRAS.509.5858H}
}

@ARTICLE{Lu2022,
       author = {{Lu}, Hong-peng and {Tian}, Hui and {Zhang}, Li-yun and {Karoff}, Christoffer and {Chen}, He-chao and {Shi}, Jian-rong and {Hou}, Zhen-yong and {Chen}, Ya-jie and {Xu}, Yu and {Wu}, Yu-chuan and {Cao}, Dong-tao and {Wang}, Jiang-tao},
        title = "{Possible detection of coronal mass ejections on late-type main-sequence stars in LAMOST medium-resolution spectra}",
      journal = {\aap},
         year = 2022,
        month = jul,
       volume = {663},
          eid = {A140},
        pages = {A140},
          doi = {10.1051/0004-6361/202142909},
archivePrefix = {arXiv},
       eprint = {2205.09972},
 primaryClass = {astro-ph.SR},
       adsurl = {https://ui.adsabs.harvard.edu/abs/2022A&A...663A.140L}
}

@ARTICLE{Berdyugina1998b,
       author = {{Berdyugina}, Svetlana V. and {Tuominen}, Ilkka},
        title = "{Permanent active longitudes and activity cycles on RS CVn stars}",
      journal = {\aap},
         year = 1998,
        month = aug,
       volume = {336},
        pages = {L25-L28},
       adsurl = {https://ui.adsabs.harvard.edu/abs/1998A&A...336L..25B}
}

@ARTICLE{Rodono2000,
       author = {{Rodon{\`o}}, M. and {Messina}, S. and {Lanza}, A.~F. and {Cutispoto}, G. and {Teriaca}, L.},
        title = "{The magnetic activity cycle of II Pegasi: results from twenty-five years of wide-band photometry}",
      journal = {\aap},
         year = 2000,
        month = jun,
       volume = {358},
        pages = {624-638},
       adsurl = {https://ui.adsabs.harvard.edu/abs/2000A&A...358..624R}
}

@ARTICLE{Leitzinger2022,
       author = {{Leitzinger}, M. and {Odert}, P.},
        title = "{Stellar Coronal Mass Ejections}",
      journal = {Serbian Astronomical Journal},
         year = 2022,
        month = dec,
       volume = {205},
        pages = {1-22},
          doi = {10.2298/SAJ2205001L},
archivePrefix = {arXiv},
       eprint = {2212.09079},
 primaryClass = {astro-ph.SR},
       adsurl = {https://ui.adsabs.harvard.edu/abs/2022SerAJ.205....1L}
}

@ARTICLE{Vida2019,
       author = {{Vida}, Kriszti{\'a}n and {Leitzinger}, Martin and {Kriskovics}, Levente and {Seli}, B{\'a}lint and {Odert}, Petra and {Kov{\'a}cs}, Orsolya Eszter and {Korhonen}, Heidi and {van Driel-Gesztelyi}, Lidia},
        title = "{The quest for stellar coronal mass ejections in late-type stars. I. Investigating Balmer-line asymmetries of single stars in Virtual Observatory data}",
      journal = {\aap},
         year = 2019,
        month = mar,
       volume = {623},
          eid = {A49},
        pages = {A49},
          doi = {10.1051/0004-6361/201834264},
archivePrefix = {arXiv},
       eprint = {1901.04229},
 primaryClass = {astro-ph.SR},
       adsurl = {https://ui.adsabs.harvard.edu/abs/2019A&A...623A..49V}
}

@ARTICLE{Aarnio2012,
       author = {{Aarnio}, Alicia N. and {Matt}, Sean P. and {Stassun}, Keivan G.},
        title = "{Mass Loss in Pre-main-sequence Stars via Coronal Mass Ejections and Implications for Angular Momentum Loss}",
      journal = {\apj},
         year = 2012,
        month = nov,
       volume = {760},
       number = {1},
          eid = {9},
        pages = {9},
          doi = {10.1088/0004-637X/760/1/9},
archivePrefix = {arXiv},
       eprint = {1209.6410},
 primaryClass = {astro-ph.SR},
       adsurl = {https://ui.adsabs.harvard.edu/abs/2012ApJ...760....9A}
}

@ARTICLE{Osten2015,
       author = {{Osten}, Rachel A. and {Wolk}, Scott J.},
        title = "{Connecting Flares and Transient Mass-loss Events in Magnetically Active Stars}",
      journal = {\apj},
         year = 2015,
        month = aug,
       volume = {809},
       number = {1},
          eid = {79},
        pages = {79},
          doi = {10.1088/0004-637X/809/1/79},
archivePrefix = {arXiv},
       eprint = {1506.04994},
 primaryClass = {astro-ph.SR},
       adsurl = {https://ui.adsabs.harvard.edu/abs/2015ApJ...809...79O}
}

@ARTICLE{Cherenkov2017,
       author = {{Cherenkov}, A. and {Bisikalo}, D. and {Fossati}, L. and {M{\"o}stl}, C.},
        title = "{The Influence of Coronal Mass Ejections on the Mass-loss Rates of Hot-Jupiters}",
      journal = {\apj},
         year = 2017,
        month = sep,
       volume = {846},
       number = {1},
          eid = {31},
        pages = {31},
          doi = {10.3847/1538-4357/aa82b2},
archivePrefix = {arXiv},
       eprint = {1709.01027},
 primaryClass = {astro-ph.EP},
       adsurl = {https://ui.adsabs.harvard.edu/abs/2017ApJ...846...31C}
}

@ARTICLE{Cohen2022,
       author = {{Cohen}, Ofer and {Alvarado-G{\'o}mez}, Juli{\'a}n D. and {Drake}, Jeremy J. and {Harbach}, Laura M. and {Garraffo}, Cecilia and {Fraschetti}, Federico},
        title = "{Space-weather-driven Variations in Ly{\ensuremath{\alpha}} Absorption Signatures of Exoplanet Atmospheric Escape: MHD Simulations and the Case of AU Mic b}",
      journal = {\apj},
         year = 2022,
        month = aug,
       volume = {934},
       number = {2},
          eid = {189},
        pages = {189},
          doi = {10.3847/1538-4357/ac78e4},
archivePrefix = {arXiv},
       eprint = {2205.08900},
 primaryClass = {astro-ph.EP},
       adsurl = {https://ui.adsabs.harvard.edu/abs/2022ApJ...934..189C}
}

@ARTICLE{Fuhrmeister2018,
       author = {{Fuhrmeister}, B. and {Czesla}, S. and {Schmitt}, J.~H.~M.~M. and {Jeffers}, S.~V. and {Caballero}, J.~A. and {Zechmeister}, M. and {Reiners}, A. and {Ribas}, I. and {Amado}, P.~J. and {Quirrenbach}, A. and {B{\'e}jar}, V.~J.~S. and {Galad{\'\i}-Enr{\'\i}quez}, D. and {Guenther}, E.~W. and {K{\"u}rster}, M. and {Montes}, D. and {Seifert}, W.},
        title = "{The CARMENES search for exoplanets around M dwarfs. Wing asymmetries of H{\ensuremath{\alpha}}, Na I D, and He I lines}",
      journal = {\aap},
         year = 2018,
        month = jul,
       volume = {615},
          eid = {A14},
        pages = {A14},
          doi = {10.1051/0004-6361/201732204},
archivePrefix = {arXiv},
       eprint = {1801.10372},
 primaryClass = {astro-ph.SR},
       adsurl = {https://ui.adsabs.harvard.edu/abs/2018A&A...615A..14F}
}

@ARTICLE{Li2017,
       author = {{Li}, Y. and {Ding}, M.~D.},
        title = "{Plasma Brightenings in a Failed Solar Filament Eruption}",
      journal = {\apj},
         year = 2017,
        month = mar,
       volume = {838},
       number = {1},
          eid = {15},
        pages = {15},
          doi = {10.3847/1538-4357/aa6348},
archivePrefix = {arXiv},
       eprint = {1702.05136},
 primaryClass = {astro-ph.SR},
       adsurl = {https://ui.adsabs.harvard.edu/abs/2017ApJ...838...15L}
}

@ARTICLE{Forbes2000,
       author = {{Forbes}, T.~G.},
        title = "{A review on the genesis of coronal mass ejections}",
      journal = {\jgr},
         year = 2000,
        month = oct,
       volume = {105},
       number = {A10},
        pages = {23153-23166},
          doi = {10.1029/2000JA000005},
       adsurl = {https://ui.adsabs.harvard.edu/abs/2000JGR...10523153F}
}

@ARTICLE{Otsu2022,
       author = {{Otsu}, Takato and {Asai}, Ayumi and {Ichimoto}, Kiyoshi and {Ishii}, Takako T. and {Namekata}, Kosuke},
        title = "{Sun-as-a-star Analyses of Various Solar Active Events Using H{\ensuremath{\alpha}} Spectral Images Taken by SMART/SDDI}",
      journal = {\apj},
         year = 2022,
        month = nov,
       volume = {939},
       number = {2},
          eid = {98},
        pages = {98},
          doi = {10.3847/1538-4357/ac9730},
archivePrefix = {arXiv},
       eprint = {2210.02819},
 primaryClass = {astro-ph.SR},
       adsurl = {https://ui.adsabs.harvard.edu/abs/2022ApJ...939...98O}
}

@article{Tei2018,
    author = {Tei, Akiko and Sakaue, Takahito and Okamoto, Takenori J and Kawate, Tomoko and Heinzel, Petr and UeNo, Satoru and Asai, Ayumi and Ichimoto, Kiyoshi and Shibata, Kazunari},
    title = "{Blue-wing enhancement of the chromospheric Mg ii h and k lines in a solar flare}",
    journal = {Publications of the Astronomical Society of Japan},
    volume = {70},
    number = {6},
    pages = {100},
    year = {2018},
    month = {05},
    issn = {0004-6264},
    doi = {10.1093/pasj/psy047},
    url = {https://doi.org/10.1093/pasj/psy047},
    eprint = {https://academic.oup.com/pasj/article-pdf/70/6/100/54674555/pasj\_70\_6\_100.pdf},
}

@ARTICLE{Koleva2021,
       author = {{Koleva}, K. and {Dechev}, M. and {Duchlev}, P.},
        title = "{Relations among eruptive prominence properties, flare evolution and CME kinematics in large solar energetic particle events}",
      journal = {Journal of Atmospheric and Solar-Terrestrial Physics},
         year = 2021,
        month = jan,
       volume = {212},
          eid = {105464},
        pages = {105464},
          doi = {10.1016/j.jastp.2020.105464},
       adsurl = {https://ui.adsabs.harvard.edu/abs/2021JASTP.21205464K}
}

@ARTICLE{Canfield1990,
       author = {{Canfield}, Richard C. and {Penn}, Matthew J. and {Wulser}, Jean-Pierre and {Kiplinger}, Alan L.},
        title = "{H alpha Spectra of Dynamic Chromospheric Processes in Five Well-observed X-Ray Flares}",
      journal = {\apj},
         year = 1990,
        month = nov,
       volume = {363},
        pages = {318},
          doi = {10.1086/169345},
       adsurl = {https://ui.adsabs.harvard.edu/abs/1990ApJ...363..318C}
}

@ARTICLE{Kennedy2015,
       author = {{Kennedy}, Michael B. and {Milligan}, Ryan O. and {Allred}, Joel C. and {Mathioudakis}, Mihalis and {Keenan}, Francis P.},
        title = "{Radiative hydrodynamic modelling and observations of the X-class solar flare on 2011 March 9}",
      journal = {\aap},
         year = 2015,
        month = jun,
       volume = {578},
          eid = {A72},
        pages = {A72},
          doi = {10.1051/0004-6361/201425144},
archivePrefix = {arXiv},
       eprint = {1504.07541},
 primaryClass = {astro-ph.SR},
       adsurl = {https://ui.adsabs.harvard.edu/abs/2015A&A...578A..72K}
}

@ARTICLE{Honda2018,
       author = {{Honda}, Satoshi and {Notsu}, Yuta and {Namekata}, Kosuke and {Notsu}, Shota and {Maehara}, Hiroyuki and {Ikuta}, Kai and {Nogami}, Daisaku and {Shibata}, Kazunari},
        title = "{Time-resolved spectroscopic observations of an M-dwarf flare star EV Lacertae during a flare}",
      journal = {\pasj},
         year = 2018,
        month = aug,
       volume = {70},
       number = {4},
          eid = {62},
        pages = {62},
          doi = {10.1093/pasj/psy055},
archivePrefix = {arXiv},
       eprint = {1804.03771},
 primaryClass = {astro-ph.SR},
       adsurl = {https://ui.adsabs.harvard.edu/abs/2018PASJ...70...62H}
}

@ARTICLE{Cao2024,
       author = {{Cao}, Dongtao and {Gu}, Shenghong},
        title = "{Red Asymmetry of H$_{ {\ensuremath{\alpha}} }$ Line Profiles during the Flares on the Active RS CVn-type Star II Pegasi}",
      journal = {\apj},
         year = 2024,
        month = mar,
       volume = {963},
       number = {1},
          eid = {13},
        pages = {13},
          doi = {10.3847/1538-4357/ad1928},
       adsurl = {https://ui.adsabs.harvard.edu/abs/2024ApJ...963...13C}
}

@ARTICLE{Namekata2024,
       author = {{Namekata}, Kosuke and {Airapetian}, Vladimir S. and {Petit}, Pascal and {Maehara}, Hiroyuki and {Ikuta}, Kai and {Inoue}, Shun and {Notsu}, Yuta and {Paudel}, Rishi R. and {Arzoumanian}, Zaven and {Avramova-Boncheva}, Antoaneta A. and {Gendreau}, Keith and {Jeffers}, Sandra V. and {Marsden}, Stephen and {Morin}, Julien and {Neiner}, Coralie and {Vidotto}, Aline A. and {Shibata}, Kazunari},
        title = "{Multiwavelength Campaign Observations of a Young Solar-type Star, EK Draconis. I. Discovery of Prominence Eruptions Associated with Superflares}",
      journal = {\apj},
         year = 2024,
        month = jan,
       volume = {961},
       number = {1},
          eid = {23},
        pages = {23},
          doi = {10.3847/1538-4357/ad0b7c},
archivePrefix = {arXiv},
       eprint = {2311.07380},
 primaryClass = {astro-ph.SR},
       adsurl = {https://ui.adsabs.harvard.edu/abs/2024ApJ...961...23N}
}

@ARTICLE{Li2021,
       author = {{Li}, Leping and {Peter}, Hardi and {Chitta}, Lakshmi Pradeep and {Song}, Hongqiang},
        title = "{On-disk Solar Coronal Condensations Facilitated by Magnetic Reconnection between Open and Closed Magnetic Structures}",
      journal = {\apj},
         year = 2021,
        month = apr,
       volume = {910},
       number = {2},
          eid = {82},
        pages = {82},
          doi = {10.3847/1538-4357/abe537},
archivePrefix = {arXiv},
       eprint = {2102.04605},
 primaryClass = {astro-ph.SR},
       adsurl = {https://ui.adsabs.harvard.edu/abs/2021ApJ...910...82L}
}
%

\end{document}